\documentclass[12pt,preprint]{aastex}

\begin{document}

\title{Correlated Infrared and X-ray Flux Changes Following the 2002
  June Outburst of the Anomalous X-ray Pulsar 1E~2259+586}

\author{Cindy R. Tam, Victoria M. Kaspi}

\affil{Department of Physics, Ernest Rutherford Physics Building,
McGill University, 3600 University Street, Montreal, QC,
H3A 2T8, Canada; tamc@physics.mcgill.ca, vkaspi@physics.mcgill.ca}

\author{Marten H. van Kerkwijk, Martin Durant}

\affil{Department of Astronomy \& Astrophysics, University of Toronto,
60 Saint George Street, Toronto, ON, M5T 1P4, Canada;
mhvk@astro.utoronto.ca, durant@astro.utoronto.ca}

\begin{abstract}
We present the results of a near-infrared monitoring program of the
Anomalous X-ray Pulsar 1E~2259+586, performed at the Gemini
Observatory. This program began three days after the pulsar's 2002
June outburst, and spans $\sim$1.5 years.  We find that after an
initial increase associated with the outburst, the near-infrared flux
decreased continually and reached the pre-burst quiescent level after
about one year. We compare both the near-infrared flux enhancement and
its decay to those of the X-ray afterglow, and find them to be
remarkably consistent.  Fitting simple power laws to the \textit{RXTE}
pulsed flux and near-infrared data for $t>1$~day post-burst, we find
the following decay indices: $\alpha =-0.21\pm0.01$ (X-ray),
$\alpha=-0.21\pm0.02$ (near-infrared), where flux is a function of
time such that $F \propto t^{\alpha}$.
This suggests that the enhanced infrared and X-ray fluxes have
a physical link post-outburst, most likely from the neutron-star
magnetosphere.  
\end{abstract}

\keywords{pulsars: individual (1E 2259+586) --- pulsars: general
  --- infrared: stars --- stars: neutron --- stars: magnetic fields}

\section{INTRODUCTION}

The origin of the optical and infrared (IR) emission in Anomalous
X-ray Pulsars (AXPs) is currently uncertain.  To date, possible
IR counterparts have been identified for 5 (of 6) known
AXPs [1E~2259+586:  \citet{htv+01,kgw+03}, 1E~1048.1$-$5937:
  \citet{wc02,ics+02}, 1RXS~J170849$-$400910: \citet{icp+03},
  4U~0142+61: \citet{hvk04} and XTE~J1810$-$197: \citet{irm+04}].  In
one case, 4U~0142+61, optical emission has been detected \citep{hvk00}
and was seen to pulse at the same period as the X-ray
pulsar \citep{km02}.  All of these sources show excess 
optical/IR emission when compared to the prediction of a simple
blackbody model extrapolated from X-ray energies (assuming the
2-component model consisting of a power-law plus blackbody component
for the X-ray emission).  The extrapolation of the power-law component
greatly overpredicts the optical/IR flux, however.

On 2002 June 18, 1E~2259+586 exhibited an outburst that included,
apart from the $>$80 bright, short-lived X-ray bursts, a large and
long-lived X-ray flux enhancement with subsequent decay
\citep{kgw+03,wkt+04,gkw04}, as well as a significant near-IR flux
enhancement, demonstrated with Gemini-North Target of Opportunity
observations made 3 and 10 days after the outburst \citep{kgw+03}.
This was the first demonstration of IR variability in an AXP, and
subsequently the first example of an associations between X-ray and IR
activity.  Since
then, IR variability has also been reported in 1E~1048.1$-$5937
\citep{wc02,ics+02,dvh04} and 4U~0142+61 \citep{hvk04} without
evidence for a correlation with X-ray outbursts, though sparsity in
X-ray observations do not preclude this.
Recently, \citet{rti+04} reported that a second near-IR observation of
the proposed counterpart to XTE~J1810$-$197 showed variablity possibly
linked to X-ray flux decay. 

In order to verify that the enhancement in IR flux seen post-outburst
in 1E~2259+586 was genuinely related to the outburst, as well as to
characterize its decay, we monitored the source using the Gemini North
telescope. Here we report on this program, demonstrating conclusively
that the IR enhancement reported by \citet{kgw+03} was associated with
the outburst.  We also find that the post-outburst IR and X-ray
radiation properties are correlated.  In \S\ref{sec:disc}, we
compare our results with expectations from different models and other
AXPs.

\section{OBSERVATIONS}

Images in the near-IR $K_s$ band ($\lambda=2.15~\mu$m, $\Delta
\lambda=0.31~\mu$m) were obtained with the Near-InfraRed Imager (NIRI;
f/6 camera; Aladdin InSb detector array; $1024\times1024$ pixels;
$0\farcs1171$ pixel size) at the 8-m Gemini North Observatory.  For a
description of the observing parameters and conditions, see
Table~\ref{tab:obs}.  The detector array was read several times in
order to reduce read-out noise.  This was done both before and after
each exposure, and the difference was recorded in the data files.  Each
target frame consisted of 4 coadded exposures of 15-s integrations.
Two earlier observations of
1E~2259+586, which took place 3 and 10 days after the X-ray burst in
June 2002 as part of a Target of Opporunity (ToO) program, were
described in \citet{kgw+03} and are also included in this analysis.
All data were reduced using the Gemini IRAF package for NIRI data.
Each frame was divided by a normalized flat field constructed from the
lamp flat frames obtained with the Gemini standard calibration unit.
The sky background image was derived
from the data frames themselves with the objects 
masked out, taking advantage of the 9-point dither pattern applied
during the observation: this was subtracted from all data frames.
Finally, all data from a single night were coadded into one image.

Photometry was performed using standard procedures within the
DAOPHOT package \citep{ste87} as implemented in IRAF.  To calibrate
the instrumental magnitudes found by DAOPHOT, we tied our measurements
directly to stars in the 1E~2259+586 field \citep{htv+01}.  The eight
nearest neighbours that were bright, isolated and not varying
\citep[stars A, B, B$'$, D, F, G, K and N under the numbering system
  of][]{htv+01} were used to measure the mean offset between the
instrumental and published $K_s$ band magnitudes.  To verify
non-variability, we chose only stars that fell within a standard
devation of $\sigma < 0.03$~mag from the weighted mean.
The advantage of this procedure was that it gave much more precise
\textit{relative} fluxes than would have been possible by using the
single standard star observed during each night.

Measured $K_s$ band magnitudes of 1E~2259+586 are listed in
Table~\ref{tab:obs}.  We analytically estimated magnitude errors
from the standard deviation of the sky background, under the
assumption that an aperture with a radius equal to the PSF FWHM
contains 70\% of a star's flux: this was added in quadrature to the
uncertainty in the photometric tie to produce final uncertainties in
Table~\ref{tab:obs}.  Our careful re-analysis of the ToO data (the
first two points) gave results consistent 
with \citet{kgw+03} well within uncertainties: the differences in
$K_s$ magnitude between the two analyses were $0.05 \pm 0.17$
(Jun. 21) and $0.18 \pm 0.25$ (Jun. 28).  The third data point was
observed with the Canada France Hawaii Telescope (CFHT) by
\citet{isc+04}: we have assumed $K'=K_s$ in our magnitude to flux
conversion.  From Table~\ref{tab:obs}, one sees that $\sim$400~days
post-burst, the source appears to have returned to its pre-burst
brightness of $K_s=21.7 \pm 0.2$~mag \citep{htv+01}.

\section{RESULTS}

The post-outburst evolution of the X-ray pulsed flux of 1E~2259+586
is described by \citet{wkt+04} in terms of a model with two power laws
in time, where $F \propto t^{\alpha}$, with $F$ the unabsorbed
2$-$10~keV pulsed flux, $t$ the time since the glitch epoch $t_g =
52443.13~$MJD, and $\alpha$ the power-law index.  Immediately after
the outburst ($<$1~day) the decay appears to follow a much steeper
power-law index 
than it does over the following year.  We compare the near-IR flux
enhancement and decay to those of the second, slower X-ray segment
consisting of flux from $>$1~day post-burst, for which \citet{wkt+04}
find a temporal decay index of $\alpha=-0.22 \pm 0.01$.

\begin{figure}
\epsscale{1.0}
\plotone{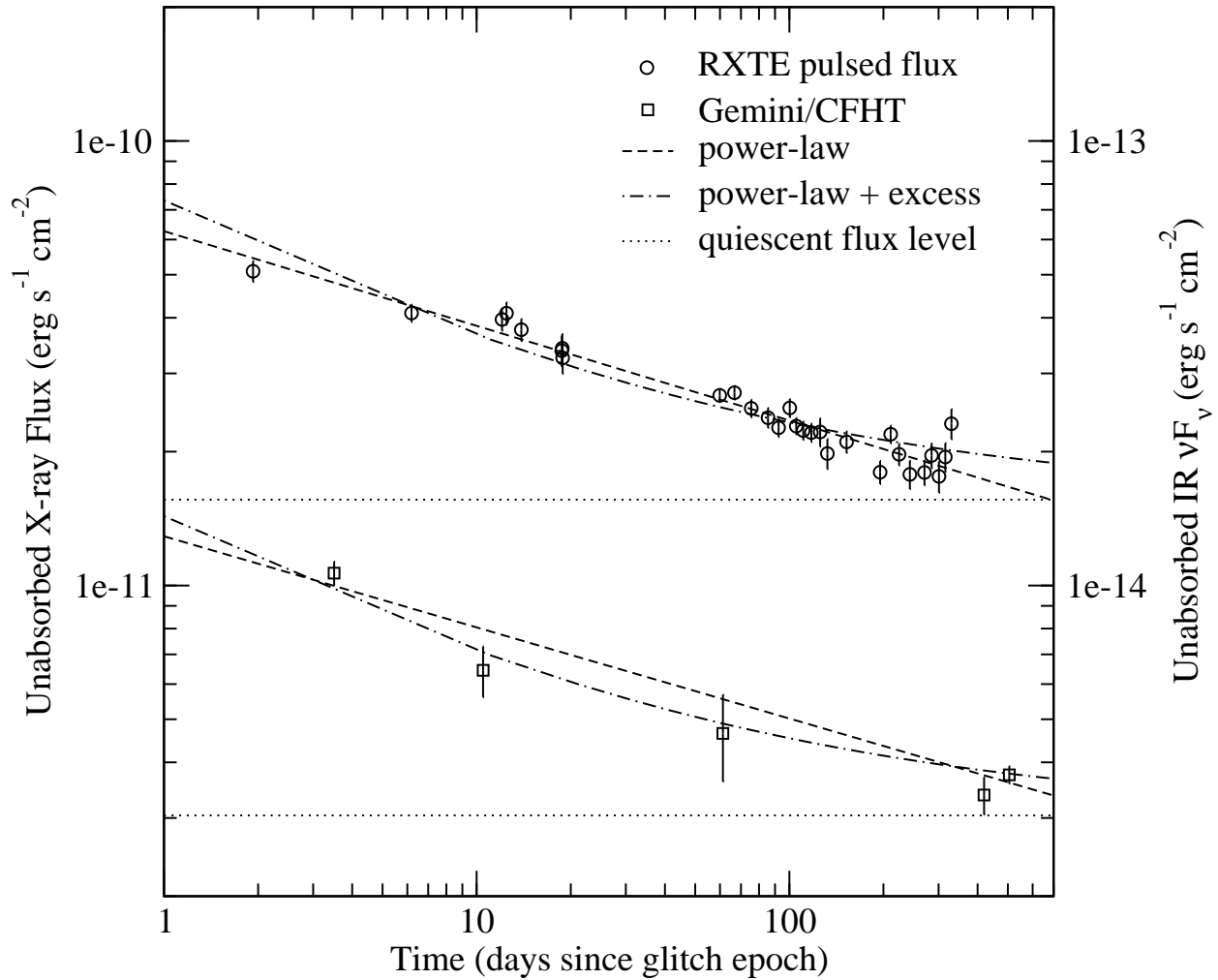}
\figcaption[f1.eps]{Unabsorbed X-ray flux and near-IR $\nu F_{\nu}$
  decay of 1E~2259+586 as a function of time.  \textit{RXTE}
  pulsed flux data \citep[from][Figure~13]{wkt+04} are represented by
  circles and refer to 
  the left axis, Gemini and CFHT data (Table~\ref{tab:obs}) are
  represented by squares and 
  refer to the right axis.  Best-fit power laws to the X-ray and
  near-IR data are shown in dashed lines.  The power-law plus excess
  model with $\alpha$ and $f$ fixed at best-fit X-ray values is shown 
  in dot-dashed lines; the dotted lines denotes the flux levels
  during quiescence as determined by the excess
  fit. \label{fig:decay}}
\end{figure}

To the near-IR data, we first apply an extinction correction of
$A_{K_s} = 0.56 \pm 0.01$, which is inferred from $A_V = N_H/(1.79
\times 10^{21}$~cm$^{-2})=5.2$~mag \citep{ps95} where $N_H=9.3 \pm 0.3
\times 10^{21}$~cm$^{-2}$ \citep{pkw+01}.  Magnitudes are then
converted to $\nu F_{\nu}$ (see Table~\ref{tab:obs} and
Figure~\ref{fig:decay}).

We fit a simple power-law function, $F=k(t/100)^{\alpha}$, to the
X-ray and IR flux using a numerical $\chi^2$ fitting program that
directly searches over parameter space, where $k$ is a constant with
dimensions erg/s/cm$^2$, $t$ is time in days since the glitch, and the
choice of the factor of 100
roughly minimizes the covariance between $k$ and $\alpha$.  Note
also that $F$ represents both X-ray flux and near-IR $\nu F_{\nu}$,
depending on the case, in erg/s/cm$^2$.  Table~\ref{tab:fit} contains
our best-fit parameter values and 1$\sigma$ uncertainties.  Fitting
the \textit{RXTE} pulsed-flux data only, we confirm the index $\alpha$ 
reported by \citet{wkt+04}\footnote{The small discrepancy can be
  explained by noting that \citet{wkt+04} perform a $\chi^2$ fit on
  log-log data to a linear function, which neglects to account for
  asymmetric uncertainties, unlike our numerical method.}. Comparing
the long-term decay of near-IR and X-ray afterglow, we find the simple
power-law indices remarkably consistent: $\alpha =-0.21 \pm 0.01$
(X-ray) and $-0.21\pm 0.02$ (IR).

Interestingly, the X-ray and IR enhancements also appear to be offset
from their respective quiescent levels $F_q$ at $t= 1$ by nearly
the same amount and, as a result, decay back to $F_q$ on similar time
scales.  To quantify this, we perform a $\chi^2$ fit to a 
second function consisting of a power-law with an excess offset
$F=F_q(1+f(t/t_0)^{\alpha})$, where $t_0=3.5$~days is the time since
the glitch of the first IR
observation, $f=(F_0 - F_q)/F_q$ is the flux excess, and $F_0$ is
the flux enhancement at $t_0$.  Data from $t<0$ are included to
determine $F_q$.  If the offsets are correlated, then we would expect
the best-fit $f$ from X-ray and $K_s$ data to be consistent: this is
in fact what we find in the latter part of Table~\ref{tab:fit}.  We
note that larger errors and a small number of near-IR data points
result in $\chi^2/\nu <1$ and large uncertainties on $f$ and $\alpha$
when all three parameters are fit to the near-IR data.  To confirm
that the shapes of the decay curves of the X-ray and $K_s$ data are
consistent statistically, we re-fit the $K_s$ data to the excess model
with $\alpha$ and $f$ fixed at the X-ray excess best-fit values.  The
effect of holding two parameters constant is a reasonable increase in
$\chi^2/\nu$ to $5.7/5 \approx 1.1$, showing that the two shapes are
indeed consistent with each other at the 1$\sigma$ level. In
Figure~\ref{fig:decay}, we plot the best-fit power laws as modelled on
the X-ray and $K_s$ data (dashed lines); overplotted is the power-law
plus excess model where $\alpha =-0.44$ and $f=2.14$ (dot-dashed
lines), and the corresponding $F_q$ best-fit value (dotted lines), for
comparison.

Thus, we find that the $>$1~day IR and X-ray initial enhancements and
subsequent decays after the 2002 outburst are correlated.

\section{DISCUSSION}
\label{sec:disc}

In this section, we compare our results with other AXPs and discuss
them in the context of various models.

AXP optical/IR emission has been argued as originating from a fossil
disk around the neutron star \citep[e.g.][]{chn00,ea03b}.  The
exceptionally high X-ray to optical/IR flux ratio seen in AXP
1E~2259+586 rendered this model problematic \citep{hvvk00,htv+01}.
Moreover,
the SGR-like bursting phenomena, observed in the 2002 outburst of 
1E~2259+586 \citep{kgw+03,gkw04,wkt+04}, and from 1E~1048.1$-$5937 in
2001 \citep{gkw02} and very recently in 2004 \citep{kgwc04}, as well
as the high optical pulsed fraction seen in AXP 4U~0142+61
\citep{km02}, simply cannot be explained by such disks. \citet{ec04}
argue that the high optical pulsed fraction can be reproduced in a
disk-dynamo model, although whether this could produce the observed
bursts is unclear.

``Hybrid'' fallback disk models, in which the disk surrounds a
magnetar, have recently been proposed to attempt to explain all AXP
properties \citep{ea03b}.  In this case, the quiescent pulsed X-ray
emission arises from accretion from the disk while optical pulsations
and bursts are magnetar magnetospheric emission.  \citet{ea03a} argue,
in the context of such a hybrid model, that enhanced X-ray emission,
following an SGR-like flare, is released from the inner disk which has
been pushed back by the burst itself.  For 1E~2259+586, however, no
such SGR-like flare was detected prior to the observed enhanced X-ray
emission, with upper limit 3 orders of magnitude below the total
observed energy release \citep{wkt+04}.  Furthermore, one AXP shows
uncorrelated torque and X-ray flux variations, contrary to the
predictions of any fallback disk model \citep{gk04}.  It is true that
one natural prediction of fallback disk models is some form of
correlation between the IR and X-ray emission \citep[see][and
  references therein]{rti+04}.  Nevertheless, in the absence of
solutions to the problematic aspects of these models as listed above,
we do not find the observed IR/X-ray correlation to render fallback
disk models particularly compelling.

By contrast, the magnetar model accounts very well for the bulk of AXP
properties, especially bursts \citep{td96a}, and qualitatively can
explain optical/IR properties as well.  In the magnetar model, thermal
surface emission is ruled out as the energy source for AXP optical/IR
emission because of the impossibly high implied brightness
temperature; hence, optical/IR emission must originate in the
stellar magnetosphere, regardless of what powers it.  Recently,
\citet{egl02} and \citet{oze04} argued that electron/positron
radiation in the magnetosphere of a magnetar, produced in analogy
with that in rotation-powered pulsars, could explain the observed
optical/IR properties of AXPs.  The strong correlation between IR and
X-ray flux decay that we have observed in the afterglow of the 2002
outburst of 1E~2259+586 argues for a physical link between the 
origins of both types of radiation.  The X-ray emission is far too
luminous \citep[$L_X \sim 10^{35}$ in 1$-$10~keV at a distance
  $\sim$4~kpc,][]{mcis02} to be rotation-powered; it follows that the
IR emission is likely not either.  Though \citet{oze04} argued that
the IR emission from an energetics standpoint could be
rotation-powered, the large implied efficiency \citep[$\nu
  F_{\nu,V}/\nu F_{\nu,rot} \sim 0.6$, from Figure~1 of][]{oze04} of
conversion from spin-down flux into IR emission in 4U~0142+61, if the
latter is rotation-powered, also argues against (though does not
disprove) this hypothesis.

The post-outburst correlation does clarify, however, the origin of the
X-ray afterglow following the 2002 outburst, because of the following
reasoning.  \citet{wkt+04} identified two possible mechanisms to
produce the X-ray afterglow.  The first was a genuine afterglow,
i.e. thermal emission from the surface, a result of an impulsive heat
injection to the crust from the magnetosphere.  Such a thermal
afterglow mechanism has been invoked in SGRs, in which the impulse,
namely a bright soft gamma-ray flare, was clearly observed
\citep{hcm+99}.  For 1E~2259+586, however, no such flare was seen
\citep{wkt+04}.  An alternative to this thermal afterglow model is
that the enhanced X-rays are a result of a twisting of the 
magnetospheric field, perhaps as a result of the twisting of its
footpoints following a signficant surface restructuring.  Such an
event is consistent with the coincidental rotational glitch that was
observed \citep{kgw+03,wkt+04}, since the latter clearly implies a
major disturbance in the crustal superfluid.  Such a twisting could
naturally result in enhanced X-rays \citep{tlk02}, with subsequent
field relaxation accounting for the decay.  Given that the IR
enhancement cannot be from surface thermal emission, the correlation
with the X-ray decay strongly favors the twisting model, as it is
difficult to understand how surface thermal X-ray emission could be so
closely correlated 
with magnetospheric radiation.  Moreover, the IR enhancement being a
result of a decaying magnetospheric disturbance is consistent with the
picture suggested by \citet{egl02} and \citet{oze04} in which the IR
emission is radiation from magnetospheric pairs.  It would be
interesting to see, in future outbursts, whether the IR emission is
pulsed and/or polarized, and if so, whether the pulse morphology is
similar, and changes in concert with any X-ray pulse morphological
changes, as this would strongly support this scenario.

IR variability over long time periods has been seen in three other
AXPs.  In the case of 1E~1048.1$-$5937, $K_s$-band variability has
been detected \citep{wc02,ics+02}.  However, no variation in X-ray
flux was seen between the different epochs.  Moreover, a more recent
observation \citep{dvh04} found that the IR flux was consistent with
the fainter of the two previous measurements, even though the X-ray
flux was significantly larger \citep{gk04}.  Despite showing no
evidence of X-ray activity \citep[and unpublished work]{gk02},
4U~0142+61 also appears to vary in $K_s$ \citep{hvk04}. This suggests
that the physical mechanism responsible for the IR emission is
distinct from that responsible for the \textit{quiescent} X-rays, even
during the broad X-ray flaring reported by \citet{gk04}. This is in
contrast to the post-outburst behavior we have seen in
1E~2259+586. That AXP IR and X-ray emission is generally correlated
following an AXP outburst is further supported by the recent report of
IR decay in data from the transient AXP candidate XTE J1810$-$197
\citep{rit+04} following its X-ray brightening and fading in 2004
\citep{ims+04}.

\acknowledgments
This work was based on observations obtained at the Gemini
Observatory (Program IDs GN-2002A-DD-6, GN-2003A-Q-71, GN-2003B-Q-22),
which is operated by the Association of Universities for Research in
Astronomy, Inc., under a cooperative agreement with the NSF on behalf
of the Gemini partnership: the National Science Foundation (United
States), the Particle Physics and Astronomy Research Council (United
Kingdom), the National Research Council (Canada), CONICYT (Chile), the
Australian Research Council (Australia), CNPq (Brazil) and CONICET
(Argentina).  It was also supported by NSERC Discovery Grant
228738-03, NSERC Steacie Supplement 268264-03, a Canada Foundation for
Innovation New Opportunities Grant, and FQRNT Team and Centre Grants.
V.~M.~K. is a Canada Research Chair and Steacie Fellow.

\bibliographystyle{apj}

\clearpage

\begin{deluxetable}{llcccccc}
\tabletypesize{\footnotesize}
\tablecaption{Near-IR observing parameters and measured results
  \label{tab:obs} }
\tablehead{
\colhead{Date} & \colhead{Instrument} & \colhead{Exposure} &
\colhead{Seeing} & \colhead{Band} & \colhead{Absorbed} &
\colhead{Unabsorbed $\nu F_{\nu}$\tablenotemark{a}}\\
\colhead{} & \colhead{} & \colhead{(min)} & \colhead{} & \colhead{} &
\colhead{magnitude} & \colhead{($10^{-15}$~erg/s/cm$^2$)} }
\startdata
2002 Jun. 21 & Gemini/NIRI & 19 & $0\farcs7$ & $K_s$ & 20.41(7) & $10.7\pm0.7$\\
2002 Jun. 28 & Gemini/NIRI & 12 & $0\farcs5$ & $K_s$ & 20.96(14) & $6.4\pm0.8$\\
2002 Aug. 18 & CFHT/AOB/KIR & 122 & $0\farcs2$ & $K'$ & 21.31(24) & $4.6\pm1.0$\\
2003 Aug. 11 & Gemini/NIRI & 58 & $0\farcs5$ & $K_s$ & 21.66(11) & $3.4\pm0.3$\\
2003 Nov. 5 & Gemini/NIRI & 49 & $0\farcs3$ & $K_s$ & 21.54(5) & $3.8\pm0.2$
\enddata
\tablenotetext{a}{$\nu F_{\nu} = 9.28 \times 10^{-7}$~erg/s/cm$^2$ for
  absorbed $K_s = 0$ \citep[Chapter~7 of][]{cox00}.}
\end{deluxetable}


\begin{deluxetable}{lllllll}
\tabletypesize{\footnotesize}
\tablecaption{Results of power-law and excess function fitting
  \label{tab:fit} }
\tablehead{
\colhead{Function} & \colhead{$F_q$ (erg/s/cm$^2$)} & \colhead{$k$
  (erg/s/cm$^2$)} & \colhead{$f$} & \colhead{$\alpha$} &
\colhead{$\chi^2$} & \colhead{$\chi^2/\nu$} } 
\startdata
X-ray PL & \nodata & $(2.35 \pm 0.02)\times 10^{-11}$ & \nodata & $-0.21 \pm 0.01$ & 37.1\tablenotemark{a} & 1.0\\
IR PL & \nodata & $(5.02 \pm 0.25)\times 10^{-15}$ & \nodata & $-0.21 \pm 0.02$ & 7.1\tablenotemark{a} & 1.0\\
X-ray excess & $(1.56 \pm 0.03)\times 10^{-11}$ & \nodata & $2.14 \pm 0.14$ & $-0.44^{+0.02}_{-0.03}$ & 97.3\tablenotemark{a} & 1.0\\
IR excess & $(3.49^{+0.22}_{-0.37})\times 10^{-15}$ & \nodata & $2.05^{+0.34}_{-0.25}$ & $-0.75^{+0.22}_{-0.33}$ & 1.4 & 0.5
\enddata
\tablenotetext{a}{Errors scaled to infer uncertainties on the
  parameters. $\chi^2$ values reflect those before rescaling.}
\end{deluxetable}

\clearpage

\end{document}